\documentclass[preprint,showpacs,showkeys,preprintnumbers,amsmath,amssymb]{revtex4}

\newcommand{\rr}{R_{\mu\nu}}

\newcommand{\rra}{2R_{{;{\mu}}{;{\nu}}}}
\newcommand{\rre}{2g_{\mu\nu}R_{;{\rho}}\,^{;{\rho}}}
\newcommand{\rri}{2R\rr-{{\frac{1}{2}g_{\mu\nu}}R^{2}}}

\newcommand{\pph}{\dot{\phi}}
\newcommand{\pphh}{\ddot{\phi}}
\newcommand{\pps}{\dot{\psi}}
\newcommand{\ppss}{\ddot{\psi}}
\newcommand{\ka}{\kappa}
\newcommand{\ds}{\displaystyle}

\usepackage{graphicx}
\usepackage{dcolumn}
\usepackage{bm}

\newcommand{\be}{\begin{equation}}
\newcommand{\en}{\end{equation}}
\newcommand{\bea}{\begin{eqnarray}}
\newcommand{\ena}{\end{eqnarray}}

\begin{document}

\preprint{GACG/0307}

\title{$R^2$- corrections to chaotic inflation}

\author{V\'{\i}ctor H. C\'ardenas}
 \email{victor.cardenas.v@mail.ucv.cl}
\author{Sergio del Campo}%
 \email{sdelcamp@ucv.cl}
\author{Ram\'{o}n Herrera}
 \email{ramon.herrera.a@mail.ucv.cl}
\affiliation{Instituto de F\'{\i}sica, Pontificia Universidad
Cat\'{o}lica de
Valpara\'{\i}so, Casilla 4059, Valparaiso, Chile.}%

\date{\today}

\begin{abstract}
Scalar density cosmological perturbations, spectral indices and
reheating in a chaotic inflationary universe model, in which a
higher derivative term is added, are investigated. This term is
supposed to play an important role in the early evolution of the
Universe, specifically at times closer to the Planck era.
\end{abstract}


\maketitle


The most appealing cosmological scenario to date is the standard
hot Big-Bang model. However, when this model is traced back to
early times in the evolution of the Universe it presents some
shortcomings (``puzzles"), such that the horizon, the flatness,
the primordial density fluctuation problems, among others.

Inflation \cite{infl} has been proposed as a good approach for
solving the cosmological ``puzzles" mentioned above. The essential
feature of any inflationary model proposed so far is the rapid but
finite period of expansion that the Universe underwent at very
early times in its evolution.

The present popularity of the inflationary scenario is entirely
due to its ability to generate a spectrum of density perturbations
which may lead to structure formation in the Universe. In this
way, it has been mentioned that inflation is a good candidate for
generating density fluctuations at scale significantly larger than
the Hubble radius without breaking causality \cite{fluct}. In
essence the conclusion that all the observations of microwave
background anisotropies performed so far \cite{COBE} support
inflation, rests on the consistency of the anisotropies with an
almost Harrison-Zel'dovich power spectrum predicted by most of the
inflationary universe models.
%
%

Among the different models we distinguish Linde's chaotic
inflationary scenario \cite{Lindcao}. This model allows various
initial distribution of the inflaton field, providing a wide class
of theories where inflation occurs under quite natural initial
conditions. In this respect, it is possible to consider a model in
which it starts at time closer to the Planck time, $t_{P} \sim
10^{-43}\, $ sec, where, it is supposed that quantum corrections
to the effective Lagrangian play an important role. It has been
suggested that the effective gravitational Lagrangian closed to
the Planck era should include higher derivative terms expressed
for instance, as a function of the scalar curvature,
$R$~\cite{Staro}. For simplicity, in this paper, we shall restrict
to the case in which the effective action presents the
$R^{2}$--term only. Generalization to include other more
complicated terms looks straightforward.

When the $R^{2}$-term of the effective action is varied with
respect to the metric tensor $g_{\mu\nu}$, it gives a term
proportional to $^{1}H_{\mu\nu}$  in the field equations. This
term coincides with that obtained as semiclassical limit of the
interaction between a quantum matter field and gravity, via the
expectation value of the corresponding energy-momentum tensor
\cite{lib1}. The tensor $^{1}H_{\mu\nu}$ is given by
$$
^{1}H_{\mu\nu}\,=\,\rra\,-\,\rre\,+\,\rri\,\,.
$$
Because higher order curvature terms are suppressed by the inverse
Planck mass squared, their effects will be significant near the
Planck era. In this respect, a higher derivative term like
$R^2$-term can be considered in addition to the Linde's chaotic
inflationary universe. This kind of model as been treated in
Refs.\cite{KLS, Gottlober}. In Ref.\cite{KLS} the authors claim
that the combined action of the $R^2$ term and the scalar field
$\phi$ may lead to Double Inflation i.e., two consecutive
inflationary stages separated by a power-law expansion. However
the analysis of perturbations made in \cite{KLS} is incomplete,
because it was assumed that both contributions can be treated
separately. These facts were elucidated in Ref.\cite{Gottlober},
where double inflation was found to occur under the following
conditions: the mass of the scalar particle has to be small
compared to the scalaron mass $m^2\ll M^2$, and the vacuum
polarization has to dominates initially. Otherwise the combined
action of both effects leads to a single inflationary phase. They
also found a break in the spectrum of perturbations (the so called
Broken Scale Invariant-BSI) for double inflation.

The aim of the present paper is to study the effects that higher
derivative term might have on inflation. Because (almost) all the
literature on this topic works in the physical frame, leading to
cumbersome expressions, we work here in the conformal frame where
all the quantities can be easily derived. Here we study the single
inflationary phase and left the double inflationary model for a
future research~\cite{future}.

Specifically, we find that inflation can be enlarged by the action
of the $R^{2}$ term. Actually, we concentrate our analysis in the
case of a single inflationary phase driven by both contributions.
We find, in agreement with previous work \cite{Gottlober}, that
this inflationary phase occurs for a special set of initial values
of the fields, constrained by the parameters of the model. We also
determine analytically the scalar density perturbations produced
during inflation valid for a non-separable effective potential, as
the one obtained in this model, which has not been discussed in
the literature. By using the observational constraint for the
amplitude of density perturbations we find that $m^2/M^2 \sim
10^{-1}$, showing consistency with a single inflationary phase. In
Ref.\cite{Gottlober} the authors use the ultrasynchroneous gauge
allowing them to treat the problem in the physical frame, where it
is found that longitudinal potentials are proportional to the
ultrasyncroneous potential. Here we will consider longitudinal
gauge quantities, which are known to be gauge
invariants~\cite{Mukanov}. Moreover, we obtain explicit analytical
expressions for the spectral indices when the universe leaves the
horizon, showing small deviations from the Harrison-Zel'dovich
spectrum, that can be observationally interesting. We also perform
in this model the analysis of reheating, based on the modern
theory of preheating.

We consider the effective action given by
\begin{eqnarray}
S=\int d^{4}x \sqrt{-g} \left[\frac{1}{2\ka^{2}} \left(R +
\frac{R^{2}}{6M^{2}} \right)+ \frac{1}{2} (\partial_{\mu}\phi)^{2}
- V(\phi) \right] \label{action1}
\end{eqnarray}
where $R$, as above, represents the scalar curvature, $\phi$ is
the ``inflaton" scalar field, $V(\phi)$ is the inflaton potential
which drives inflation, and $M$ is an arbitrary parameter with
dimension of Planck mass, $m_{P}\,\sim\,10^{19}\,GeV$ and
$\ka^{2}\,=\,8\pi/m^{2}_{P}$.

In order to calculate the scalar density perturbation we take the
following conformal transformation \cite{Whitt}
\begin{equation}
\bar{g}_{\mu\nu}\,=\,\Omega^{2}(x)\,g_{\mu\nu}, \label{eq2}
\end{equation}
with
\begin{equation}
\Omega^{2}(x)\,=\,( 1\,+\,\frac{R}{3\,M^{2}}),\label{eq3}
\end{equation}
where $g_{\mu\nu}$ and $\bar{g}_{\mu\nu}$ represent the original
and the transformed metric, respectively. Introducing a new field
$\psi$ defined by
\begin{equation}
\psi=\frac{1}{\ka\beta}ln(1+\frac{R}{3M^{2}}),\label{eq4}
\end{equation}
where $\beta\,\equiv\,\sqrt{2/3}$, and substituting eqs.
(\ref{eq2}-\ref{eq4}) in the action (\ref{action1}), it  becomes a
Hilbert-Einstein-type action given by
\begin{eqnarray}
\bar{S}= \int d^{4}x \sqrt{-\bar{g}}
\{\frac{\bar{R}}{2\ka^{2}}+\frac{1}{2}(\partial\psi)^{2} +
\frac{1}{2}e^{-\beta\ka\psi}(\partial\phi)^2+\nonumber\\
-\widetilde{V}(\phi,\psi)\},\label{action2}
\end{eqnarray}
where the potential $\widetilde{V}(\phi,\psi)$ takes the form
\begin{equation}
\widetilde{V}(\phi,\psi)=\frac{3M^{2}}{4\,\ka^{2}}(1-e^{-\beta\ka\psi})^{2}\,+\,
e^{-2\beta\ka\psi}\,V(\phi). \label{pot}
\end{equation}
Notice that, if we disregard the inflaton field $\phi$, then
inflation is described entirely by gravity, since the model
becomes supported by the $R^{2}$-theory \cite{r2}.

Let us introduce into the action (\ref{action2}) the conformal
Friedmann-Robertson-Walker (FRW) metric \be
d\,\bar{s}^{2}\,=\,d\,\bar{t}^{2}\,-\,
\bar{a}\,(\,\bar{t}\,)^{2}\,dx_{i}dx^{i}, \en where $dx_{i}dx^{i}$
and $\bar{a}(\,\bar{t}\,)$ represent the flat three-surface and
the conformal scale factor, respectively. Considering also that
the corresponding scalar fields $\phi$ and $\psi$ are homogeneous,
i.e., $\phi=\phi(\bar{t})$ and $\psi(\bar{t})$, then, after
variations of this action respect to the different field
variables, the following Equations of motion are found
$$
3\,\bar{H}\,^{2}\,=\,\ka^{2}\left[\frac{1}{2}\dot{\psi}^{2}\,+\,\frac{1}{2}e^
{-\beta\ka\psi}\dot{\phi}^{2}\,+\,\widetilde{V}(\psi,\phi)\right],
$$
\be \ppss\,
+\,3\,\bar{H}\,\pps\,=\,-\frac{\beta\,\ka}{2}\,e^{-\beta\ka\psi}
\dot{\phi}^{2}\,-
\,\frac{\partial{\widetilde{V}(\psi,\phi)}}{\partial{\psi}},
\label{Eqsconf} \en and
$$
\pphh\, +\,3\,\bar{H}\,\pph\,=\,\beta\,\ka\,\pps\,\pph-
\,e^{\beta\ka\psi}\frac{\partial{\widetilde{V}(\psi,\phi)}}{\partial{\phi}},
$$
where the dots denote differentiation respect to the time
$\bar{t}$ and $\bar{H}\,=\,\dot{\bar{a}}/\bar{a}$ defines the
Hubble parameter in the conformal frame.

In the slow-roll over approximation the field eqs. (\ref{Eqsconf})
become,
$$
3\,\bar{H}\,^{2}\,=\,\ka^{2}\,\widetilde{V}(\psi,\phi),
$$
\be
3\,\bar{H}\,\pps\,=\,-\,\frac{\partial{\widetilde{V}(\psi,\phi)}}{\partial{\psi}},
\label{Eqsconf2} \en and
$$
3\,\bar{H}\,\pph\,=\,-\,e^{\beta\ka\psi}\frac{\partial{\widetilde{V}(\psi,\phi)}}
{\partial{\phi}}.
$$
This set of eqs. is valid if the following conditions are met
$$
max\{\dot{\psi}^{2}\,,\,e^{-\beta\ka\psi}\dot{\phi}^{2}\}\,\ll\,\widetilde{V}(\psi,\phi),
$$
$$
e^{\beta\ka\psi}\,(\widetilde{V}(\psi,\phi)_{,\phi})^{2}\,\ll\,\widetilde{V}(\psi,\phi)
\mid \widetilde{V}(\psi,\phi)_{,\psi}\mid,
$$
and
$$ e^{-\beta\ka\psi}\,\widetilde{V}(\psi,\phi)_{,\phi\phi}\,\ll\, \mid
\widetilde{V}(\psi,\phi)_{,\psi}\mid,
$$
where $\widetilde{V}(\psi,\phi)_{,\phi}$ and
$\widetilde{V}(\psi,\phi)_{,\psi}$ represent the derivatives of
the potential $\widetilde{V}(\psi,\phi)$ respect to the fields
$\phi$ and $\psi$, respectively. A solution to the equation of
motion (\ref{Eqsconf2}) for the chaotic inflationary potential, $
V(\phi)=\frac{1}{2}m^{2}\phi^{2}$\cite{Lindcao} and
$M^2$$\neq2m^2$ is given by
 \be \label{app}
e^{\beta\ka\psi}\,=\,\frac{\beta^2\ka^2\,m^2}{(M^2-2m^2)}\,\phi^2+1\approx\,
\frac{\beta^2\ka^2\,m^2}{(M^2-2m^2)}\,\phi^2,
 \en
where, we have taken the integration constant
$C=0\Longrightarrow\,e^{\beta\ka\psi_{o}}\,-1=\,
\frac{\beta^2\ka^2\,m^2}{(M^2-2m^2)}\,\phi^2_{o}$. The subscript
${0}$ denotes the values of each field at the beginning of
inflation. The approximation involve in (\ref{app}) is used for
simplicity. In this case, we find
 \be
 \phi^{2}\,=\,\phi_{o}^{2}\,-\frac{(M^2-2m^2)}{\bar{H}_o\ka^{2}}\,\bar{t}\,
 \en
 and
\be \psi\,=\,\frac{1}{\beta\ka}\,ln\left[\,e^{\beta\ka\psi_{o}}\,
-\frac{2\,m^{2}}{3\,\bar{H}_o}\,\bar{t} \right],
 \en
where
 $$
\bar{H}_o^{2}=\frac{M^{2}}{4}(\,1\,-\,e^{-\beta\ka\psi_o}\,)^{2}
+\frac{m^{2}\ka^2\phi^2_o}{6}\,e^{-2\beta\ka\psi_o}=\frac{\ka^2\widetilde{V_o}}{3}.
 $$

Since we are mainly interested in the existence of inflationary
stage, we should take the case in which $M >\sqrt{2}m$. Here, the
shape of the potential exhibits an appropriated form for inflation
to occur.  This is not the case if $M \leq \sqrt{2}m$, since the
potential presents an unacceptable peak for certain values of the
scalar fields, which would force us to restrict strongly the range
of these fields for inflationary models to occur.

The period of the superluminar expansion can be determined by the
value of the inflaton field $\phi$ where the slow-rollover
approximation breaks down. This happens when
$|\ddot{\phi}|\simeq|3\bar{H}\dot{\phi}|$. This condition gives
\be
\phi^2(\bar{t}_{e})\,=\phi_{e}^2\simeq\,\frac{(M^2-2m^2)}{6\bar{H}_o^2\ka^2},
 \en
and the inflationary period results in \be
\bar{t}_{e}\,\simeq\,\frac{\bar{H}_o\ka^2\,\phi_o^2}
{(M^2-2m^2)}\,\,-\,\frac{1}{6\bar{H}_o},
 \en
where the subscript $e$ denotes the values at the end of
inflation.

The requirement for solving some of the puzzles present in
standard Big-Bang cosmology is \be
 \frac{a(t_e)}{a(0)}=\frac{\bar{a}(\bar{t}_e)}{\bar{a}(0)}\,\,
e^{\beta\ka/2(\psi_o-\psi(\bar{t}_e)}\gtrsim\,e^{65}.
  \en
This results in a condition for the initial value of the inflaton
field given by \be
 9\,\beta^2\,\bar{H}_o^2\,\phi_o^2\,\ka^2\,\gtrsim\,8^2\,[M^2-2m^2]
 \en
or
 \be
 \ka^2\,\phi^2_o\,\gtrsim\,\,\frac{\,[\,M^2-2m^2\,]}{M^2}
 \left[46-\frac{M^2}{\beta^2m^2} \right],
 \en
which dependents on the parameters $M$ and $m$, as we could expect
for two-field models with a no separable potential.

In order to calculate the scalar density fluctuations, we follow a
similar approach to that described in ref.~ \cite{Star2}. We take
the perturbed metric in the longitudinal gauge, which is known to
be manifestly gauge-invariant \cite{Mukanov}. Therefore, in
comoving coordinates we write \be
d\,\bar{s}\,^{2}\,=\,(1\,+\,2\bar{\Phi})\,d\,\bar{t}\,^{2}\,
-\,\bar{a}\,^{2}(1\,-2\bar{\Psi})\,\delta_{ij}\,dx^{i}\,dx^{j},
\label{metpert} \en where the scalar metric perturbation fields
are functions of the space and time coordinates, i.e.,
$\bar{\Phi}=\bar{\Phi}(\vec{x},\bar{t})$ and $\bar{\Psi}=
\bar{\Psi}(\vec{x},\bar{t})$.

In order to describe a linear theory for the cosmological
perturbations, we introduce fluctuations into the scalar fields
$\phi$ and $\psi$, so that we write
$\phi(\vec{x},\overline{t})=\phi(\overline{t})+
\delta\phi(\vec{x},\overline{t})$ and
$\psi(\vec{x},\overline{t})=\psi(\overline{t})+
\delta\psi(\vec{x},\overline{t})$, where the background fields,
$\phi(\overline{t})$ and $\psi(\overline{t})$, are solutions to
the homogeneous Einstein equations, and the deltas  are small
perturbations that represent small fluctuations of the
corresponding scalar fields.

Since the spatial part of the resultant energy-momentum tensor
becomes diagonal, then it found that $\bar{\Phi}=\bar{\Psi}$. With
this result, we can write the perturbed Einstein field equations,
in which we can take  spatial Fourier transform for the different
variables. It is found that each mode satisfies the corresponding
linearized perturbed field equation, given by
\begin{eqnarray}
\dot{\bar{\Phi}}+\bar{H}
\bar{\Phi}=\frac{\ka^{2}}{2}\left[e^{-\beta\ka\psi}\dot{\phi}
\delta\phi+\dot{\psi}\delta\psi \right],
\end{eqnarray}
\begin{eqnarray}
\delta\ddot{\phi}+(3\bar{H}-\beta\ka\dot{\psi})
\delta\dot{\phi}+(\frac{k^2}{\bar{a}^2}+e^{\beta\ka\psi}
\widetilde{V}_{,\,\phi\phi})\delta\phi \nonumber\\
-\beta\ka\dot{\phi}\delta\dot{\psi}+(e^{\beta\ka\psi}
\widetilde{V}_{,\,\phi})_{,\,\psi}\delta\psi=
\nonumber\\
2(\ddot{\phi}+3\bar{H}\dot{\phi})\bar{\Phi}+4\dot{\phi}\dot{\bar{\Phi}}-
2\beta\ka\dot{\phi}\dot{\psi}\bar{\Phi},
\end{eqnarray}

and
\begin{eqnarray*}
\delta\ddot{\psi}+3\bar{H}\delta\dot{\psi}+(\frac{k^{2}}
{\bar{a}^{2}}-\frac{\beta^{2}\ka^{2}}{2}
e^{-\beta\ka\psi}\dot{\phi}^{2}+\widetilde{V}_{,\,\psi\psi})\delta\psi
\\
+\beta\ka e^{-\beta\ka\psi}\dot{\phi} \delta\dot{\phi} +
\widetilde{V}_{,\,\psi\phi}\delta\phi
\end{eqnarray*}
\vspace{-0.3cm}
\begin{eqnarray}
=2(\ddot{\psi}+3\bar{H}\dot{\psi})\bar{\Phi}+4\dot{\psi}\dot{\bar{\Phi}}+\beta\ka\,e^{-\beta\ka\psi}\dot{\phi}^{2}\bar{\Phi}\,.
 \end{eqnarray}
If in this set of equations we neglect terms proportional to
$\dot{\Phi}$ and two time derivatives, and we consider
non-decreasing adiabatic and isocurvature modes on large scale,
i.e. $k\ll\overline{a}\,\overline{H}$, the following equations are
obtained~ \cite{Staroper} \be
\bar{\Phi}\,=\,\frac{\ka^{2}}{2\overline{H}}\,\left[\,e^{-\beta\ka\psi}\,\dot{\phi}\,
\delta\phi\,+\,\dot{\psi}\,\delta\psi\, \right], \en
\be
 3\bar{H}\delta\dot{\phi}+(e^{\beta\ka\psi}\widetilde{V}_{,\,\phi})_{,\,\psi}\delta\psi+
 e^{\beta\ka\psi}\widetilde{V}_{,\,\phi\phi}\,
 \delta\phi+2e^{\beta\ka\psi}\widetilde{V}_{,\,\phi}\bar{\Phi}=0,
\en and
\be
 3\bar{H}\delta\dot{\psi}+\widetilde{V}_{,\,\psi\psi}\delta\psi+
 \widetilde{V}_{,\,\psi\phi}\delta\phi+2V_{,\,\psi}\bar{\Phi}\,,
\en where $\widetilde{V}$ is the potential specified by
eq.~(\ref{pot}).

For $V(\phi)=m^2\phi^2/2$ we can solve the above equations,
following a similar approach to ref.~ \cite{Staroper}. It is found
that
 \be\label{ecpert1}
 \frac{\delta\phi}{\dot{\phi}}\,=\,\frac{1}{\bar{H}}\,
 [\,C_{1}+C_{2}-C_{2}\,f_{1}(\psi,\phi)\,]\,,\en
 \be
 \frac{\delta\psi}{\dot{\psi}}\,=\,\frac{1}{\bar{H}}\,
 [\,C_{1}+C_{2}\,f_{2}(\psi,\phi)\,]\,\label{ecpert2}\en
 and
 \be
 \bar{\Phi}\,=\,-C_{1}\,\frac{\dot{\bar{H}}}{\bar{H}^2}+\frac{C_{2}}{\ka^{2}}
 \left(f_{2}\frac{\widetilde{V}_{,\,\psi}^2}{V^2}+\frac{\widetilde{V}_{,\,\phi}^2}
 {\widetilde{V}^2}e^{\beta\ka\psi}(1-f_{1})\right),\label{ecpert3}\en
where
$$
f_{1}=f_{1}(\psi,\phi)=\alpha_1(\phi_e,\psi_e)+e^{-\beta\ka\psi}
$$
$$
 \hspace{-.05cm}\times \left[C_{e2}\left(\frac{3M^2}{4\ka^2}
(2-e^{-\beta\ka\psi})- \frac{m^2\phi^2}{2}e^{-\beta\ka\psi}
\right)-2C_{e1} \right],
 $$
and
\begin{widetext}
$$
f_2=f_{2}(\psi,\phi)=\alpha_2(\phi_e,\psi_e)
+\left[\frac{3M^2}{\left(3M^2(e^{-\beta\ka\psi}-1)
+2\ka^2m^2\phi^2e^{-\beta\ka\psi}\right)}\right]
$$
$$
\times\left[C_{e2}\left(\frac{m^2}{2}\phi^2e^{-
\beta\ka\psi}-\widetilde{V}\right)\ln\left(\frac{3M^2(e^{-\beta\ka\psi}-1)}
 {4\ka^2}+\frac{m^2\phi^2e^{-\beta\ka\psi}}{2}
\right)+C_{e1}(e^{-\beta\ka\psi}-1)\right]
$$
\end{widetext}
where $\alpha_1$ and $\alpha_2$ are arbitrary constants and
$C_{e1}$ and $C_{e2}$ are given by $\ds
C_{e1}=\frac{e^{-\beta\ka\psi_e}}{\widetilde{V_e}}
\left[\frac{3M^2}{4\ka^2}(1-e^{\beta\ka\psi_e})+\frac{m^2\phi_{e}^2}{2}
\right]$ and $\ds C_{e2}=\frac{1}{\widetilde{V_e}} $,
respectively.

In the  expressions (\ref{ecpert1}-\ref{ecpert3}), terms in
proportion to $C_{1}$ and $C_{3}$ represent adiabatic and
isocurvature modes, respectively.

>From eqs.~ (\ref{ecpert1}) and (\ref{ecpert2}) the curvature
perturbation due to primordially adiabatic fluctuation at the end
of inflation,  when the scale crosses the horizon is given by
 \be
 \frac{\delta\bar{\rho}}{\bar{\rho}}\,\approx\,\bar{H}_h\left[\,\frac{\delta\psi}{\dot{\psi}}
 [\,1\,-\,f(\psi,\phi)\,]\,+\,\frac{\delta\phi}{\dot{\phi}}\,f(\psi,\phi)\,
 \right]_{h}\,,
\en where
$$
f(\psi,\phi)\,=\,\frac{f_{2}}{[\,f_{1}\,+\,f_{2}\,-\,1\,]}.
$$
In the limit $M^{-1}\longrightarrow 0\Longrightarrow \psi =0$,
which gives $f=Cte. $ and thus we obtain the standard result for
the density perturbation
$\delta\bar{\rho}/\bar{\rho}\propto\bar{H}^2/\dot{\phi}$. The
magnitude of the density perturbations take the form
 $$
\hspace{-4.0cm}\frac{\delta\bar{\rho}}{\bar{\rho}}
\approx\frac{\beta\kappa^{3}\bar{H}_o^{3}\phi_h^{2}}{[M^{2}-2m^{2}]}
$$
\be
\hspace{2.0cm}\times\left[1+f_{\phi_{h}}\left(\frac{2m}{\sqrt{(M^2-2m^2)}}-1\right)
\right],\label{delta}
 \en
 where $\phi_{h}$ represent the field scalar $\phi$ at witch the universe
 crosses the Hubble radius and $f_{\phi_{h}}$ is the function $f$ evaluate
 at $\phi_{h}$. Since
 $\bar{a}(\bar{t_h})/\bar{a}(\bar{t_e})=k\bar{H}_{o}^{-1}$ we find
 \be
 \phi_h^2=\frac{(M^2-2m^2)}{\bar{H}_o^2\kappa^2}\left[\ln(\bar{H}_ok^{-1})+\frac{1}{6}
\right],\label{phih}
 \en
and in view  that the log term is of the order of ${\cal{O}}(10
^{2})$ and $\delta\bar{\rho}/\bar{\rho}\sim 10^{-5}$, we find,
from eqs.~ (\ref{delta}) and~ (\ref{phih}), the following order
for the parameters of our model $m^2/M^2\sim 10^{-1}$. For this
order the initial value of $\phi$ becomes restricted from below
$\kappa\phi_o\gtrsim 5$.

Another interesting quantity to investigate is the spectral index
$n_{S}$ of the power spectrum of primordial fluctuations, which
becomes specified by $P(k)\propto\,k^{n_{S}}$. It is well known
that most of the inflationary scenarios are compatible with an
almost Harrison-Zel'dovich spectrum, i.e., $n_{S}\,\approx\,1$,
and thus, it provides detailed information about the early time in
the evolution of the universe. This prediction of the inflationary
universe models is sustained by the COBE data~ \cite{COBE}, which
are statistically compatible with the $n_{S}\approx\,1$ result.

In our model, we could obtain the spectral index by using the
following standard expression
\begin{eqnarray}
\bar{n}_{S}-1=\,\frac{d\,\ln\,\mid\, \frac{\delta\,
\bar{\rho}}{\bar{\rho}}\,\mid^{2}} {d\,\ln\,k},
\end{eqnarray}
where, together with eqs.~(\ref{delta}) and~(\ref{phih}), it is
obtained,
$$
\bar{n}_S-1=\frac{-2}{[\ln(\bar{H}_o^2k^{-1})+\frac{1}{6}]}+
f_k'\left[\frac{(f_k-1)+\frac{m^2}{(M^2-2m^2)}f_k}{(1-f_k)^2+
\frac{m^2}{(M^2-2m^2)}f_k^2}\right]
$$
where $f_k$ represents the function $f$ expressed in terms of $k$
and $f'_{k}=d\,f_{k}/dln k$.
The right-hand-size of the latter equation represents the
deviation from the $\bar{n}_{S}=1$ Harrison--Zel'dovich spectrum.
For instant, if we use the constraint obtained aboved, i.e.,
$m^2/M^2\sim10^{-1}$, we obtaine that $\bar{n}_{S}\approx0.98$,
which differs from the Harrison--Zel'dovich (HZ) value
approximately in a $2\%$. We should note that this deviation is
inside of the observational range, since it has been reported that
the scalar perturbations do not differ from scale invariant by a
large amount, since $|n_{S}\,-\,1|\,\sim\,{\cal{O}}\,(0.1)$, i.e.,
inside of the 10$\%$ \cite{HuTu}.

In addition to the scalar curvature perturbations there are tensor
or gravitational wave perturbations. These perturbations can also
be generated from quantum fluctuations during inflation. Since we
have chosen to work in the Einstein conformal frame we can use the
standard results for the evolution of tensor perturbations. The
spectrum of the gravitational wave $P_G$ is known to be given by
 $P_{G}=8\ka^2(\bar{H}/2\pi)^2$, and
where the spectral index  is defined by
$\bar{n}_{G}=dlnP_{G}/dlnk$~\cite{Mukanov,Liddle}. Thus, we find
in our case
$$
\bar{n}_G=\left[\frac{3M^4}{4\kappa^4\bar{H}_o^2\beta}
\ln(\bar{H}_ok^{-1})-1\right]^{-1}\,
\left[\ln(\bar{H}_ok^{-1})+\frac{1}{6}\right]^{-1}
$$
where $\beta=9M^2(M^2-2m^2)/(8\ka^4m^2)$. We could give an order
of magnitude of this parameter if we use the value obtained for
the quotient $m/M$ determined above. Our result gives
$\bar{n}_{G}\approx-0.22$. Notice that this value differs in about
a 20$\%$ of that value predicted by the HZ spectrum which gives
$\bar{n}_{G}\sim 0$. Thus, our model presents a significant
deviation from the HZ result. Only astronomical observations will
tell us which spectrum will be the right one.


Note that every expression was determined in the Einstein
conformal frame. In order to see what happens with the same
expressions but in the physical frame, we have to use the original
conformal transformation eqs.~(\ref{eq2}) and (\ref{eq3}). The
quantities in the Einstein frame are related to the physical frame
by means of a conformal transformation given by eq.(\ref{eq2}). In
both frame, the gauge-invariant perturbations are related by
\cite{Hwang} \be
\Phi\,=\,\bar{\Phi}\,-\,\frac{\delta\Omega}{\Omega}\,\,, \en where
from eqs. (3) and (4) it is found that
$\delta\Omega\,=\,\beta\,\Omega\,\delta\psi/2$, with $
\delta\psi\,\sim\,\bar{H}$.

In the physical frame, the gravitational potential becomes \be
\Phi\,=\,C_{1}\left(\,1\,-\,\left[H\,+\,2\frac{\Omega_{,t}}{\Omega}\right]\,
\frac{1}{a\Omega^{2}}\,\int^{t}\,a\Omega^{2}\,dt'\right)\,. \en
Comparing both expressions, it is observed that they are very
different, but, during inflation, where the Hubble parameter
remains practically constant, we expect that the quadratic scalar
curvature term in the effective Lagrangian varies slowly, so that
the $\Omega$ conformal factor could be considered constant during
this period. Therefore, the adiabatic fluctuations become
described by the same expression in both frames. The same argument
can be draw for the spectral indices $\bar{n}_S$ and $\bar{n}_G$,
since it depends directly of the expression for
$\delta\bar{\rho}/\bar{\rho}$ and $\bar{H}$ respectively.

Every inflationary model needs a mechanism to reheat the universe,
i.e., a physical mechanism capable to patch the inflationary era
with the radiation dominated phase (needed for example in
nucleosynthesis). The current understanding of this process
comprises three phases: preheating (the resonant amplification
phase), the perturbative decay and thermalization \cite{reheat}.
Reheating in the $R^{2}$ model has been considered by Suen and
Anderson~\cite{rehR2} by solving the semiclassical backreaction
equations. Recently Tsujikawa et {\it{al}}.~\cite{tsuj} have
studied the preheating phase with nonminimally coupled scalar
fields. However, as far as we know the study of preheating in the
model that we are considering here, has not been discussed in the
literature.

In order to study the preheating phase we consider the inflaton
field coupled with another scalar field, say $\chi$ through the
interaction term
\begin{equation}
-\frac{1}{2}g^{2}\phi^{2}\chi^{2}, \label{int}
\end{equation}
where $g$ is the coupling constant. During the coherent
oscillations of the inflaton field around the minimum at the
origin, which is precisely the period where reheating occur, the
scalaron field moves towards the minimum, but in a time scale
shorter than that of the inflaton $\phi$. In essence we can
consider it nearly constant during preheating; which is the first
part of reheating. Under these assumptions the inflaton field
equation reads
\begin{equation}
\ddot{\phi} +
3\bar{H}\dot{\phi}=-e^{\beta\ka\psi}(m^{2}+g^{2}\langle\chi^{2}\rangle)\phi,
\label{phie}
\end{equation}
where we have replaced $\chi^{2}$ by the quantum expectation
value,  $\langle\chi^{2}\rangle$, doing manifest the mean field
approximation involved. The precise definition of this object will
be given in short. Furthermore, the equation of the $\chi$ field
is expressed by
\begin{equation}
\ddot{\chi}+3\bar{H}\dot{\chi}-\nabla^{2}\chi
=-e^{\beta\ka\psi}(g^{2}\phi^{2})\chi. \label{chieq}
\end{equation}
As we can notice here, the preheating phase proceed very similar
to the standard way, because the only difference is the
exponential damping term in the coupling.
%
%
%
By considering the field $\chi$ as a quantum operator we can
expand it as
\begin{equation}
\chi=\int \frac{d^{3}k}{(2\pi)^{3/2}}\left[ a_{k}\chi_{k}(\bar{t})e^{-i\textbf{kx}}+%
a_{k}^{\dag}\chi_{k}^{*}(\bar{t})e^{i\textbf{kx}}\right],
\end{equation}
where the $a$'s are annihilation and creation operators, and the
mode functions $\chi_{k}(t)$ satisfy the equation
\begin{equation}
\ddot{\chi_{k}}+3\bar{H}\dot{\chi_{k}}+\left[\left(\frac{k}{\bar{a}}\right)^{2}%
+e^{\beta\ka\psi}g^{2}\phi^{2}\right]\chi_{k}=0.\label{modeeq}
\end{equation}
The expectation value of $\chi^{2}$ is expressed in terms of the
mode functions as
\begin{equation}
\langle\chi^{2}\rangle=\frac{1}{2\pi^{2}}\int dk
k^{2}|\chi_{k}^{2}|.
\end{equation}
Eq.~(\ref{modeeq}) can be rewritten as a harmonic oscillator with
a time dependent frequency
\begin{equation}
\ddot{X_{k}}+\Omega_{k}^{2}(\bar{t})X_{k}=0,\label{haosc}
\end{equation}
where we have defined the variable
$X_{k}=\bar{a}(\bar{t})^{2/3}\chi_{k}$ and the frequency is given
by
\begin{equation}
\Omega_{k}^{2}=\left(\frac{k}{\bar{a}}\right)^{2}+e^{\beta\ka\psi}g^{2}\phi^{2}%
-\frac{3}{4}\bar{H}^{2}-\frac{3}{2}\frac{\ddot{\bar{a}}}{\bar{a}}.
\label{freq}
\end{equation}
Usually the last two terms in (\ref{freq}) are of the same order
and also very small compare with the effective mass $m_{\chi}$, so
we can neglect these terms as a first approximation. In this case
the frequency depends on the scale factor $\bar{a}(\bar{t})$, the
scalar field $\phi(\bar{t})$ and $\psi(\bar{t})$.

During the oscillations of the inflaton field $\phi$, the scalaron
$\psi$ first rolls towards the minimum and after that start to
oscillate around it, right after a few oscillations of $\phi$.
Preheating occurs here before $\psi$ start to oscillate. If we
define $\overline{\psi}_{0}$ as the value at this point, an
approximated solution of (\ref{phie}) is
\begin{equation}
\phi(\bar{t})\simeq \Phi(\bar{t}) sin(\omega \bar{t}),
\end{equation}
where $\omega=e^{\beta \kappa \overline{\psi}_{0}}m$ and the
amplitude $\Phi(\bar{t})$ changes slowly with $\bar{t}$. Inserting
this solution in eq.~(\ref{freq}) we find a Mathieu type equation.
It controls both the narrow and broad resonance particle
production. We have performed numerical studies of that equation
finding no surprises, just the evident effect of damping terms in
(\ref{phie}) and (\ref{haosc}). After $\psi$ start to oscillate
(for $\psi<\overline{\psi}_{0}$), the system becomes completely
dominated by $\psi$, implying another period of reheating, this
time without any resonance amplification, in which the $\psi$
field decays. A detailed research of preheating in these models,
considering both inflationary realizations, is under study
\cite{future}.

\begin{acknowledgments}
VHC was supported from COMISION NACIONAL DE CIENCIAS Y TECNOLOGIA
through FONDECYT Project 3010017 grant. SdC was also supported
from COMISION NACIONAL DE CIENCIAS Y TECNOLOGIA through FONDECYT
Projects N$^0$ 1030469 and 1010485 grants, and from UCV-DGIP N$^0$
123.764/2003. RH is supported from MINISTERIO DE EDUCACION through
MECESUP Projects FSM 9901 and USA 0108 grants.

\end{acknowledgments}


\end{document}